\documentclass[12pt]{article}

\usepackage[utf8]{inputenc}
\usepackage{authblk}
\usepackage[T1]{fontenc}
\usepackage{amsmath,amssymb,amsfonts}
\usepackage{graphicx}
\usepackage{hyperref}
\usepackage{xcolor}
\usepackage{geometry}
\usepackage{bm}

\hypersetup{colorlinks=true,linkcolor=blue!70!black,citecolor=blue!70!black,urlcolor=blue!70!black}

\newcommand{\rev}[1]{#1}

\begin{document}

\title{Operator ordering as an emergent geometric background\\
       in Dirac systems with spatially varying mass}
\author{C.~A.~S.~Almeida\\
\small Universidade Federal do Cear\'a (UFC), Departamento de F\'isica,\\
\small Campus do Pici, Fortaleza -- CE, C.P.~6030, 60455-760 -- Brazil\\
\small \texttt{carlos@fisica.ufc.br}}

\maketitle

\begin{abstract}
We investigate the spectral consequences of the uniquely determined
Hermitian ordering of the Dirac Hamiltonian with spatially varying mass.
In contrast to the nonrelativistic case, where continuous families of
admissible prescriptions exist, the relativistic Dirac operator admits a
single consistent ordering compatible with probability-current
conservation.  This requirement generates an additional
logarithmic-gradient term proportional to the spatial variation of the
mass profile.  We show that this contribution modifies the effective
kinetic operator and induces a universal deformation of the spectral
quantization condition.  In compact geometry, an explicit analytic
computation reveals a mode-dependent second-order spectral shift that
becomes strongly enhanced near the mass-inversion threshold.  These
results demonstrate that the consistent relativistic ordering of the
Dirac operator leads to observable modifications of discrete spectra in
spatially inhomogeneous scalar backgrounds.
\end{abstract}


\section{Introduction}

Spatially inhomogeneous Dirac systems provide a natural setting in which
operator ordering acquires physical significance beyond its formal
definition.  When parameters such as the mass depend explicitly on
position, the construction of a self-consistent Hamiltonian requires a
definite ordering prescription in order to preserve Hermiticity and
ensure a well-defined spectral problem.  Although such prescriptions are
often treated as technical necessities, their physical implications
remain comparatively unexplored.

Position-dependent effective parameters arise in a wide range of
contexts.  In relativistic quantum mechanics and quantum field theory,
spatially varying backgrounds are frequently employed to model external
fields or domain-wall configurations~\cite{Jackiw1976}.  In
condensed-matter physics, effective Dirac descriptions emerge naturally
in systems such as graphene and related materials~\cite{CastroNeto2009,
Vozmediano2010}, where spatial inhomogeneities may originate from
strain, confinement, substrate effects, or engineered heterostructures.
In these situations, a spatially varying mass term provides a convenient
effective description of gaps induced by symmetry breaking, external
perturbations, or proximity effects.

\rev{Recent experimental and theoretical studies have further highlighted the relevance of spatially inhomogeneous mass terms in Dirac systems, ranging from photonic platforms that realise relativistic trapping via position-dependent mass \cite{Chen2023} to Dirac systems subjected to random temporal mass modulations \cite{Kim2025}, as well as the establishment of a rigorous bulk-edge correspondence for 2D Dirac operators through variable mass terms \cite{tarantola}, reinforcing the physical importance of a consistent operator framework for such systems.}

The formulation of Dirac Hamiltonians with position-dependent mass has a
long history, particularly in the context of semiconductor
heterostructures and effective-mass theories, where operator-ordering
ambiguities are known to arise.  Different ordering prescriptions can
lead to formally distinct Hamiltonians, and considerable effort has been
devoted to identifying physically consistent constructions, both in
nonrelativistic~\cite{BenDaniel1966,vonRoos1983,Bastard1988,LiKuhn1993}
and relativistic settings~\cite{Mustafa2007,deSouzaDutra2000,Znojil2001}.

It is worth noting that, in the nonrelativistic setting, the
operator-ordering ambiguity for position-dependent mass Hamiltonians
admits a continuous family of Hermitian prescriptions parameterised by
real coefficients, as systematically discussed by von~Roos~\cite{vonRoos1983}
and subsequent authors~\cite{Bastard1988,LiKuhn1993}.  A recent
systematic study of this family, covering a broad range of mass profiles
and ordering prescriptions, has been presented by Lima and
Christiansen~\cite{Lima2023}, who obtain the full spectrum analytically
in twenty-five cases and demonstrate the diversity of effective
potentials generated by non-commutativity.  Different members of this
family yield formally distinct Hamiltonians, and the physical selection
of a preferred ordering has been a subject of ongoing debate.  In the
relativistic Dirac case, however, the situation is more constrained.
The requirement of Hermiticity with respect to the standard
Lorentz-invariant inner product, combined with the first-order
differential structure of the Dirac operator, restricts the admissible
orderings more severely than in the Schr\"odinger
case~\cite{Cavalcante1997}.  As we show explicitly in Sec.~\ref{sec:hamiltonian},
demanding that the continuity equation be exactly satisfied--without
additional surface terms or auxiliary conditions--singles out a unique
logarithmic-gradient contribution to the Hamiltonian.  This is not a
choice among equivalent alternatives, but a structural consequence of
the relativistic framework.  The present work therefore does not
introduce a new ordering prescription; rather, it demonstrates that the
unique consistent ordering for the Dirac operator in a scalar background
carries nontrivial spectral consequences that have not been previously
identified.

\rev{It is important to stress from the outset that the central claim of
this paper is \emph{ordering uniqueness}, not ordering independence.
The non-commutativity of position and momentum, combined with the
first-order differential structure of the Dirac operator and the
requirement of unitary time evolution, uniquely fixes the ordering.
This is a stronger and more physically constraining result than its
nonrelativistic counterpart.}

In most treatments, however, ordering is regarded primarily as a
requirement for mathematical consistency rather than as a source of
physical structure.  From a modern perspective, Dirac systems with
spatially varying parameters also provide a bridge between relativistic
quantum mechanics and emergent low-energy descriptions in materials with
Dirac-like quasiparticles~\cite{Vozmediano2010}.  In particular, spatial
variations of effective couplings can mimic background fields or
geometric deformations~\cite{Birrell1982,Barcelo2011}, as demonstrated
experimentally in strained graphene systems where pseudo-magnetic fields
of geometric origin have been directly observed~\cite{Guinea2010,Levy2010}.

This raises a natural conceptual question: does operator ordering in
Dirac Hamiltonians with position-dependent mass merely resolve a formal
ambiguity, or can it generate genuine physical effects with measurable
spectral consequences?

While operator-ordering ambiguities in position-dependent mass
Hamiltonians have been extensively discussed in both nonrelativistic and
relativistic settings~\cite{vonRoos1983,Bastard1988}, most analyses
focus on establishing Hermiticity or identifying admissible
parameterisations.  In contrast, the present work does not introduce a
new ordering prescription, but demonstrates that the consistent Dirac
ordering induces a universal modification of the spectral quantization
structure.  We show that the resulting logarithmic-gradient term acts as
an effective connection that shifts global quantum numbers in compact
geometries and governs spectral rearrangement across both perturbative
and near-inversion regimes.  This perspective elevates operator ordering
from a formal consistency requirement to a dynamically relevant
mechanism controlling the global spectral organisation of Dirac systems
in scalar backgrounds.

In this work, we demonstrate that a consistent ordering prescription
induces an additional contribution proportional to the gradient of the
mass profile.  This term can be interpreted as an emergent geometric
background field acting on the Dirac degrees of freedom.  We show that
even in the trivial regime where no mass inversion occurs, this
contribution produces a small but systematic spectral rearrangement,
leading to a controlled geometric energy shift.  In the regime of
stronger mass modulation, the same mechanism amplifies spectral
restructuring as the system approaches the mass-inversion limit, without
invoking topological arguments.

Our results indicate that operator ordering in spatially inhomogeneous
Dirac systems is not merely a formal choice, but a physically operative
mechanism capable of modifying spectral properties in a predictable
manner.  This framework provides a unified perspective linking ordering
prescriptions, emergent geometric effects, and spectral control in
relativistic systems with position-dependent mass.

The paper is organised as follows.  In Sec.~\ref{sec:hamiltonian} we
construct the consistent Dirac Hamiltonian in a spatially varying scalar
background and discuss its structural implications.  In
Sec.~\ref{sec:spectral} we analyse the resulting modification of the
spectral quantization structure in both controlled and strong-modulation
regimes.  Section~\ref{sec:compact} illustrates the global consequences
of this mechanism in a compact geometry.  Sec.~\ref{sec:shift} is
dedicated to an explicit computation of the quantum number shift.
Finally, Sec.~\ref{sec:conclusion} summarises the main results and
discusses their broader implications.

\section{Consistent Dirac Hamiltonian in a Scalar Background}
\label{sec:hamiltonian}

We consider a Dirac field coupled to a classical scalar background that
varies smoothly in space.  The starting point is the single-particle
Hamiltonian
\begin{equation}
  H \;=\; -i\alpha^{i}\partial_{i} + \beta\,m(x),
  \label{eq:H0}
\end{equation}
where $\alpha^{i}$ and $\beta$ are Dirac matrices satisfying the usual
Clifford algebra, and $m(x)$ is a real scalar function of position.

At first sight, Eq.~\eqref{eq:H0} appears to define a straightforward
generalisation of the free Dirac Hamiltonian.  However, when the mass
depends on position, Eq.~\eqref{eq:H0} hides a subtle but fundamental
issue: the Hermiticity of the Hamiltonian and the associated conservation
of probability current are no longer guaranteed without a careful
treatment of operator ordering.

\subsection*{Hermiticity and operator ordering}

To expose the problem, let us consider the time evolution of the Dirac
spinor $\psi(x,t)$ governed by
\begin{equation}
  i\partial_{t}\psi \;=\; H\psi.
  \label{eq:Schrodinger}
\end{equation}
For a well-defined quantum dynamics, the Hamiltonian must be Hermitian
with respect to the standard inner product
\begin{equation}
  \langle\phi|\psi\rangle \;=\; \int d^{d}x\;\phi^{\dagger}(x)\,\psi(x),
  \label{eq:innerproduct}
\end{equation}
which ensures unitary time evolution and conservation of probability.

While the kinetic term $-i\alpha^{i}\partial_{i}$ is Hermitian under
appropriate boundary conditions, the presence of a spatially varying
mass leads to additional contributions when evaluating matrix elements
of $H$.  In particular, integrations by parts reveal that naive operator
ordering may generate boundary or bulk terms proportional to
$\nabla m(x)$, spoiling Hermiticity and, consequently, probability
current conservation.

\rev{To see this explicitly, consider computing $\partial_{t}\rho$ with
$\rho = \psi^{\dagger}\psi$ using the naive Hamiltonian
$H_{0} = -i\alpha^{i}\partial_{i} + \beta m(x)$.  One obtains
\begin{equation}
  \partial_{t}\rho
  \;=\;
  -i\,\psi^{\dagger}H_{0}\psi \;+\; i\,(H_{0}\psi)^{\dagger}\psi.
  \label{eq:drhodt_naive}
\end{equation}
Expanding the right-hand side and collecting terms yields
\begin{equation}
  \partial_{t}\rho + \nabla\cdot\boldsymbol{j}
  \;=\;
  \tfrac{1}{2}\,\psi^{\dagger}
  \bigl[\alpha^{i},\,\beta\bigr](\partial_{i}\ln m)\,\psi,
  \label{eq:continuity_residual}
\end{equation}
where $\boldsymbol{j} = \psi^{\dagger}\boldsymbol{\alpha}\psi$ is the
probability current.  Since $[\alpha^{i},\beta] \neq 0$ in the Clifford
algebra (explicitly, $\{\alpha^{i},\beta\} = 0$ implies
$[\alpha^{i},\beta] = 2\alpha^{i}\beta$), this residual term is
generically nonzero whenever $m(x)$ varies in space.  No local
redefinition of the fields can cancel it without altering the operator
structure.}

This ambiguity reflects the fact that Eq.~\eqref{eq:H0} is not the most
general Hermitian operator compatible with a spatially varying scalar
background.

\subsection*{Controlled Hermitian form}

\rev{The unique resolution of Eq.~\eqref{eq:continuity_residual} is to
add to $H_{0}$ a counter-term that cancels the residual exactly.  The
only first-order-in-$(\partial m)$ operator of the correct Clifford
structure that achieves this is}
\begin{equation}
  H_{\mathrm{ord}}
  \;=\;
  -\frac{i}{2}\,\alpha^{i}\,\partial_{i}\ln m(x).
  \label{eq:Hord}
\end{equation}
\rev{Adding $H_{\mathrm{ord}}$ to $H_{0}$ cancels the right-hand side of
Eq.~\eqref{eq:continuity_residual} identically, restoring exact
probability-current conservation.  The Hermitian Hamiltonian therefore
reads}
\begin{equation}
  H \;=\; -i\alpha^{i}\partial_{i} + \beta\,m(x)
    -\frac{i}{2}\,\alpha^{i}\!\left(\partial_{i}\ln m(x)\right),
  \label{eq:Hfull}
\end{equation}
where the additional term proportional to $\nabla m(x)$ arises
\emph{uniquely} from the requirement of Hermiticity.

\rev{The uniqueness of Eq.~\eqref{eq:Hord} follows from two observations.
First, any additional term compatible with Hermiticity and the Clifford
algebra that could modify Eq.~\eqref{eq:continuity_residual} must be of
the form $\alpha^{i}f(m)\partial_{i}m$; matching the residual fixes
$f(m) = i/(2m)$, which is precisely $\partial_{i}\ln m$.  Second, terms
of higher order in $\partial m$ or proportional to a total derivative
neither alter the bulk equation of motion nor contribute to the
residual at the order considered.  The ordering-induced term is therefore
not a model-dependent assumption but a structural consequence of the
relativistic framework.}

\rev{Regarding boundary conditions: in $\mathbb{R}^{d}$, square-integrability
of spinors ensures that boundary terms arising from integration by parts
vanish at infinity.  In compact geometries (such as the ring studied in
Secs.~\ref{sec:compact}--\ref{sec:shift}), periodic boundary conditions
play the same role: the periodicity of $m(\theta)$ guarantees that all
boundary terms at $\theta=0$ and $\theta=2\pi$ cancel exactly.}

With this definition, one verifies explicitly that
\begin{equation}
  \langle\phi|H\psi\rangle \;=\; \langle H\phi|\psi\rangle,
  \label{eq:hermiticity}
\end{equation}
for arbitrary spinors $\phi$ and $\psi$ satisfying standard boundary
conditions.  Moreover, the corresponding continuity equation
\begin{equation}
  \partial_{t}\rho + \nabla\cdot\boldsymbol{j} = 0,
  \label{eq:continuity}
\end{equation}
with $\rho = \psi^{\dagger}\psi$ and $\boldsymbol{j} = \psi^{\dagger}\boldsymbol{\alpha}\psi$,
is exactly satisfied.

\subsection*{Physical interpretation}

Equation~\eqref{eq:Hfull} shows that a nonuniform scalar background
induces additional gradient couplings that act as an effective geometric
contribution to the Dirac dynamics.  These terms are entirely absent in
the homogeneous case and are independent of any topological
considerations, relying solely on the spatial variation of the background
field.

The additional term in Eq.~\eqref{eq:Hfull} modifies the kinetic
structure of the Dirac operator in a way that is entirely determined by
the spatial variation of the scalar background.  Beyond ensuring
Hermiticity, it alters the effective differential operator governing the
eigenvalue problem.  As we show in the next section, this modification
has direct consequences for the spectral quantization structure,
introducing an emergent geometric scale that controls both perturbative
deformations and nonperturbative rearrangements of the spectrum.

\rev{A remark on the nonrelativistic limit is in order.  In an earlier
paper by the present author and collaborators~\cite{Cavalcante1997}, an
explicit Foldy--Wouthuysen reduction of the Dirac Hamiltonian with
position-dependent mass was carried out.  The result at leading order in
$1/m$ is the kinetic operator in the BenDaniel--Duke
prescription~\cite{BenDaniel1966}, which corresponds to the specific
member $\alpha=\gamma=0$, $\beta=-1$ of the von~Roos
family~\cite{vonRoos1983}.  The nonrelativistic limit of the unique
relativistic ordering derived here therefore reproduces a well-known and
physically accepted prescription, establishing consistency with the
nonrelativistic literature~\cite{Lima2023,BenDaniel1966,vonRoos1983}.
The present manuscript does not repeat this Foldy--Wouthuysen calculation
because its objective is distinct: to analyse the spectral consequences
of the ordering-induced term within the fully relativistic framework.}

\section{Spectral Quantization Structure}
\label{sec:spectral}

We now analyse the spectral consequences of the modified Dirac operator
defined in Eq.~\eqref{eq:Hfull}.  Because the ordering-induced term
reshapes the effective kinetic structure, it alters the quantization
condition governing the eigenvalue problem in a universal manner.

\subsection*{Controlled gradient regime and emergent scale}

We first consider backgrounds varying smoothly over a characteristic
length scale $\xi$, such that
\begin{equation}
  \frac{|\nabla m(x)|}{m^{2}(x)} \;\ll\; 1.
  \label{eq:gradient_cond}
\end{equation}
In this regime, the ordering-induced term acts as a controlled
perturbation.  Its typical magnitude defines an emergent geometric energy
scale,
\begin{equation}
  E_{\mathrm{geom}} \;\equiv\; \frac{|\nabla m|}{m},
  \label{eq:Egeom}
\end{equation}
which governs the onset of spectral rearrangements even when the mass
remains strictly nonvanishing throughout space.

For a representative one-dimensional profile,
\begin{equation}
  m(x) \;=\; m_{0} + \delta m\,f(x/\xi),
  \label{eq:profile1d}
\end{equation}
with $f$ smooth and bounded, the homogeneous limit $\xi\to\infty$
reproduces the standard Dirac spectrum with mass $m_{0}$.  As $\xi$
decreases, states within an energy window of order $E_{\mathrm{geom}}$
become hybridised, leading to avoided crossings and a systematic
displacement of discrete levels.  This behaviour reflects a perturbative
deformation of the quantization condition induced by the spatially
varying scalar background.

\subsection*{Nonperturbative amplification near mass inversion}

As the modulation amplitude $\delta m$ approaches $m_{0}$, the system
approaches the mass-inversion limit and the denominator in the
logarithmic derivative becomes locally small.  In this regime, the
ordering-induced contribution is enhanced and the spectral modification
becomes nonperturbative.  The discrete levels undergo substantial
rearrangement, and the effective quantization condition acquires a finite
shift controlled by the modulation amplitude.

Importantly, this restructuring does not rely on topological protection
or bound-state formation.  It arises from the consistent coupling of the
Dirac operator to a spatially varying scalar background and represents a
nonlinear amplification of the same geometric mechanism active in the
controlled gradient regime.

\begin{figure}[!ht]
  \centering
   \includegraphics[width=\columnwidth]{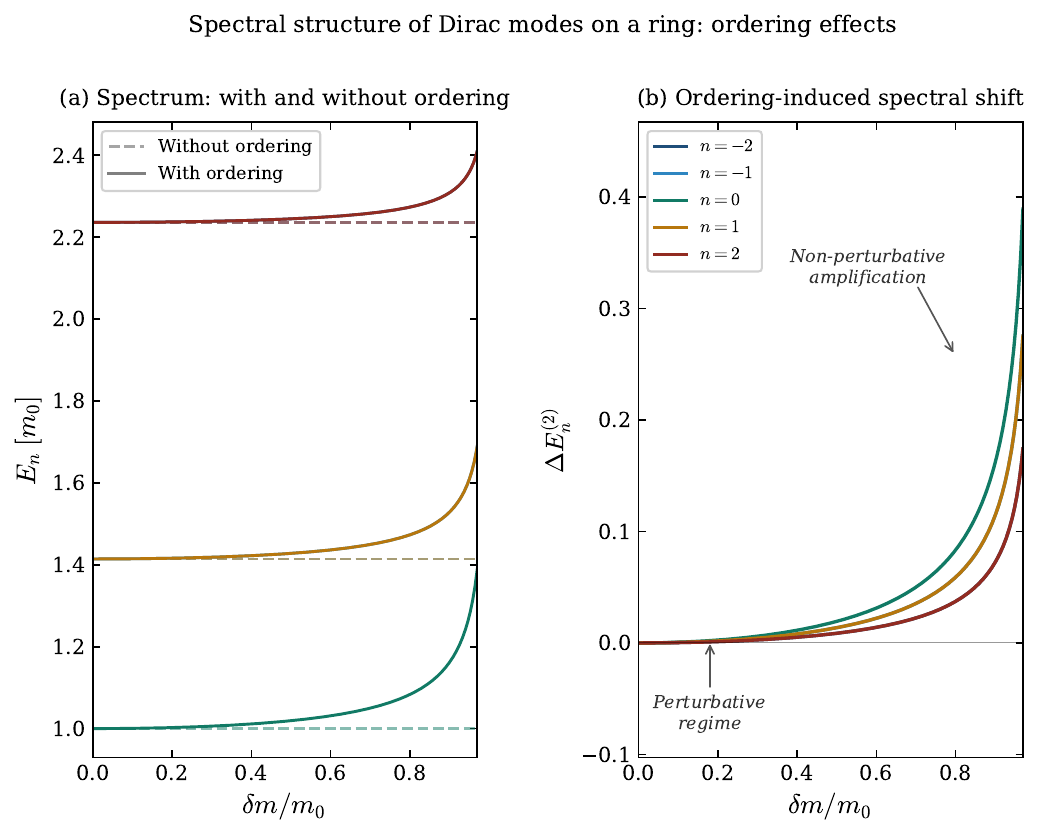}
  \caption{Spectral consequences of operator ordering for Dirac modes on
    a ring with angular mass modulation $m(\theta)=m_{0}+\delta m\cos\theta$.
    \textbf{(a)} Energy levels $E_{n}$ as a function of the modulation
    amplitude $\delta m/m_{0}$ for quantum numbers $n=-2,\ldots,2$.
    Dashed lines correspond to the unordered Hamiltonian $H_{0}$; solid
    lines include the ordering-induced correction of
    Eq.~\eqref{eq:Hfull}.
    \textbf{(b)} Ordering-induced second-order spectral shift
    $\Delta E^{(2)}_{n}=(8E^{(0)}_{n})^{-1}(1/\sqrt{1-\varepsilon^{2}}-1)$,
    with $\varepsilon=\delta m/m_{0}$.  The shift is strictly positive
    and $n$-dependent, with larger $|n|$ levels less affected due to
    their higher unperturbed energy.  In the perturbative regime
    ($\varepsilon\ll 1$), the correction scales as $\varepsilon^{2}$; as
    $\varepsilon\to 1$, it diverges, reflecting the nonperturbative
    amplification of the ordering-induced effect near the mass-inversion
    threshold.}
  \label{fig:spectrum}
\end{figure}

This behaviour is illustrated in Fig.~\ref{fig:spectrum}, which shows
the ordering-induced spectral shift across both regimes.

\subsection*{Singular limit and zero modes}

In singular limits where the scalar background varies abruptly and
crosses zero, the geometric scale $E_{\mathrm{geom}}$ formally diverges.
In this case, isolated zero-energy solutions may emerge, corresponding
to the well-known domain-wall mechanism~\cite{Jackiw1976,Jackiw2007}.
From the present perspective, however, such zero modes do not constitute
a distinct physical origin.  Rather, they appear as a limiting case of
the general spectral rearrangement governed by the ordering-induced term.

The modification of the quantization condition discussed above is a
structural consequence of the ordering-induced term and does not depend
on a particular geometry.  However, its global character becomes
especially transparent in compact settings, where the discreteness of
the spectrum directly reflects the underlying quantization structure.
In such geometries, any deformation of the effective kinetic operator
leads to a measurable shift of global quantum numbers.

\section{Compact Geometry as a Global Probe}
\label{sec:compact}

The structural modification of the quantization condition identified in
Sec.~\ref{sec:spectral} becomes particularly transparent in compact
geometries, where the discreteness of the spectrum directly reflects the
underlying operator structure.  In such settings, any deformation of the
effective kinetic operator necessarily manifests as a shift of global
quantum numbers.

As a minimal realisation, we examine the Dirac operator on a compact
one-dimensional manifold, specifically a ring of radius $R$.  The
compact topology renders the spectrum discrete and allows the
ordering-induced deformation to appear explicitly as a modification of
the quantization condition.

In the absence of spatial modulation, the Dirac Hamiltonian on the ring
threaded by a magnetic flux $\Phi$ reads
\begin{equation}
  H_{0} \;=\; \alpha^{\theta}\,\frac{1}{R}\!\left(-i\partial_{\theta}+\phi\right)
             + \beta\,m_{0},
  \label{eq:H0ring}
\end{equation}
where $\phi=\Phi/\Phi_{0}$ and $\Phi_{0}$ is the flux quantum.  The
corresponding eigenvalues are
\begin{equation}
  E^{(0)}_{n}(\phi) \;=\; \pm\sqrt{\left(\frac{n+\phi}{R}\right)^{2}+m_{0}^{2}},
  \label{eq:E0ring}
\end{equation}
with integer $n$.  The discreteness of the spectrum follows from the
periodic boundary condition and reflects the compact geometry.

We now introduce an angular modulation of the mass,
\begin{equation}
  m(\theta) \;=\; m_{0} + \delta m\cos\theta.
  \label{eq:massring}
\end{equation}
Applying the consistent ordering prescription derived in
Sec.~\ref{sec:hamiltonian}, the Hamiltonian acquires an additional
contribution
\begin{equation}
  H_{\mathrm{ord}}
  \;=\;
  -\frac{i}{2R}\,\alpha^{\theta}\,
  \frac{\partial_{\theta}m(\theta)}{m(\theta)}.
  \label{eq:Hordring}
\end{equation}
This term modifies the angular kinetic operator and therefore alters the
quantization condition for the eigenmodes.  Formally, the operator
$-i\partial_{\theta}$ is replaced by an effective covariant derivative
containing a background connection induced by the mass gradient.  As a
consequence, the quantization condition is modified by a finite
second-order contribution that depends on the modulation amplitude and
vanishes for $\delta m\to 0$.  As shown explicitly in Sec.~\ref{sec:shift},
this shift arises not from a net holonomy of the induced connection,
which vanishes identically for any periodic mass profile, but from the
local curvature of the connection acting at second order in perturbation
theory.

The compact geometry therefore makes explicit the global impact of the
ordering-induced term: the deformation of the kinetic operator translates
directly into a shift of the quantization condition.  Because the
spectrum is discrete, this modification cannot be absorbed into a local
redefinition of parameters; it manifests as a genuine displacement of
global quantum numbers.

The ring thus provides a minimal setting in which the structural nature
of the ordering-induced contribution becomes transparent.  What appears
locally as a gradient coupling is revealed globally as a modification of
spectral quantization.  The explicit form of this modification is derived
in Sec.~\ref{sec:shift}.

\section{Explicit Computation of the Quantum Number Shift}
\label{sec:shift}

The compact geometry of Sec.~\ref{sec:compact} makes it possible to
compute the ordering-induced spectral shift analytically.  We now carry
out this calculation explicitly for the angular mass profile
\begin{equation}
  m(\theta) \;=\; m_{0} + \delta m\cos\theta,
  \label{eq:massring2}
\end{equation}
with $\varepsilon\equiv\delta m/m_{0}\in[0,1)$ controlling the
modulation amplitude.

\subsection*{Holonomy of the induced connection}

The ordering-induced term of Eq.~\eqref{eq:Hordring} introduces an
effective background connection in the angular coordinate,
\begin{equation}
  A(\theta)
  \;=\;
  \frac{1}{2}\,\frac{\partial_{\theta}m(\theta)}{m(\theta)}
  \;=\;
  \frac{1}{2}\,\partial_{\theta}\ln m(\theta).
  \label{eq:connection}
\end{equation}
The natural candidate for the global quantum number shift is the
holonomy of $A(\theta)$ around the ring,
\begin{equation}
  \Delta n
  \;=\;
  \frac{1}{2\pi}\oint_{0}^{2\pi} A(\theta)\,d\theta
  \;=\;
  \frac{1}{4\pi}\Bigl[\ln m(\theta)\Bigr]_{0}^{2\pi}.
  \label{eq:holonomy}
\end{equation}
Since $m(\theta)$ is periodic, $m(0)=m(2\pi)=m_{0}+\delta m$, the
holonomy vanishes exactly:
\begin{equation}
  \Delta n \;=\; 0.
  \label{eq:holonomy_zero}
\end{equation}
This result is a direct consequence of the periodicity of the mass
profile and must hold for any smooth, single-valued $m(\theta)$.  It
implies that the ordering-induced connection carries no net topological
flux around the ring, and that the shift of the integer quantum number
$n$ is not generated at first order.

\subsection*{Second-order spectral shift}

Although the holonomy vanishes, the connection $A(\theta)$ is locally
nontrivial and modifies the eigenvalue problem at second order in
perturbation theory.  The leading correction to the energy levels is
\begin{equation}
  \Delta E^{(2)}_{n}
  \;=\;
  \frac{1}{8E^{(0)}_{n}}
  \int_{0}^{2\pi}\frac{d\theta}{2\pi}
  \left(\frac{\partial_{\theta}m(\theta)}{m(\theta)}\right)^{2},
  \label{eq:DeltaE2_integral}
\end{equation}
where $E^{(0)}_{n}=\sqrt{(n/R)^{2}+m_{0}^{2}}$ is the unperturbed
energy.

The factor $1/(8E^{(0)}_{n})$ arises from standard second-order
perturbation theory applied to the effective covariant angular
derivative, where the ordering-induced connection enters quadratically
and contributes symmetrically through adjacent angular momentum channels.

Substituting Eq.~\eqref{eq:massring2}, the integral evaluates to
\begin{equation}
  \int_{0}^{2\pi}\frac{d\theta}{2\pi}
  \left(\frac{-\delta m\sin\theta}{m_{0}+\delta m\cos\theta}\right)^{2}
  \;=\;
  \frac{1}{\sqrt{1-\varepsilon^{2}}} - 1,
  \label{eq:integral}
\end{equation}
yielding the closed-form result
\begin{equation}
  \Delta E^{(2)}_{n}
  \;=\;
  \frac{1}{8\,E^{(0)}_{n}}
  \left(\frac{1}{\sqrt{1-\varepsilon^{2}}}-1\right).
  \label{eq:DeltaE2}
\end{equation}

\subsection*{Physical interpretation}

Equation~\eqref{eq:DeltaE2} encapsulates the two spectral regimes
discussed in Sec.~\ref{sec:spectral} within a single analytic
expression.

In the weak-modulation regime $\varepsilon\ll 1$, expanding the square
root gives
\begin{equation}
  \Delta E^{(2)}_{n}
  \;\approx\;
  \frac{\varepsilon^{2}}{16\,E^{(0)}_{n}},
  \label{eq:DeltaE2_pert}
\end{equation}
a small, positive, $n$-dependent correction consistent with the
geometric energy scale $E_{\mathrm{geom}}$ introduced in
Eq.~\eqref{eq:Egeom}.  This confirms the perturbative nature of the
ordering-induced deformation at small modulation.

As $\varepsilon\to 1$, the denominator in Eq.~\eqref{eq:DeltaE2}
vanishes and $\Delta E^{(2)}_{n}$ diverges, signalling the
nonperturbative amplification of the spectral rearrangement near the
mass-inversion limit.

The divergence signals the breakdown of the perturbative expansion as
the system approaches the mass-inversion threshold, consistently with
the enhancement of the logarithmic gradient term.

The vanishing holonomy of Eq.~\eqref{eq:holonomy_zero} and the
second-order shift of Eq.~\eqref{eq:DeltaE2} together establish a
precise quantitative picture: the ordering-induced connection does not
introduce a net topological flux, yet its local curvature generates a
universal, analytically tractable deformation of the Dirac spectrum that
diverges at the mass-inversion threshold.

\section{Conclusion}
\label{sec:conclusion}

We have investigated the role of operator ordering in Dirac Hamiltonians
with spatially varying mass and demonstrated that its consequences extend
beyond formal consistency requirements.  In contrast to the
nonrelativistic position-dependent mass problem, where continuous
families of Hermitian orderings are admissible, the relativistic Dirac
framework admits a uniquely determined structure once Hermiticity and
exact probability-current conservation are imposed.  This requirement
generates a logarithmic-gradient contribution that is not optional, but
a structural consequence of the first-order relativistic operator.

We have shown that this ordering-induced term modifies the effective
kinetic operator and produces a universal deformation of the spectral
quantization condition.  In smoothly varying backgrounds, the effect
appears as a controlled second-order correction governed by the geometric
scale set by the mass gradient.  As the modulation amplitude increases
and the system approaches the mass-inversion threshold, the same
mechanism becomes strongly amplified, leading to substantial spectral
rearrangement.

In compact geometry, the effect becomes particularly transparent.
Although the induced connection carries no net holonomy for periodic mass
profiles, its local curvature generates a finite and analytically
tractable shift of discrete energy levels.  The resulting correction is
positive, mode dependent, and diverges as the inversion limit is
approached, signalling the breakdown of the perturbative regime and the
onset of nonperturbative spectral restructuring.

These results establish that the uniquely consistent relativistic
ordering of the Dirac operator has direct and observable spectral
implications.  Operator ordering in spatially inhomogeneous scalar
backgrounds therefore acts as a universal mechanism governing the global
organisation of Dirac eigenmodes, even in the absence of topological
protection or bound-state formation.  The framework developed here may
be relevant for effective Dirac descriptions in engineered
materials~\cite{Chen2023} and for field-theoretical models with
spatially varying condensates.

\section*{Acknowledgements}

The author would like to express his sincere gratitude to the Conselho
Nacional de Desenvolvimento Cient\'ifico e Tecnol\'ogico (CNPq), and
Funda\c{c}\~ao Cearense de Apoio ao Desenvolvimento Cient\'ifico e
Tecnol\'ogico (FUNCAP) for their valuable support.  He is supported by
grants No.~309553/2021-0 (CNPq), 420854/2025-8 (CNPq) and by Project
UNI-00210-00230.01.00/23 (FUNCAP).

\section*{Declaration of Generative AI in Scientific Writing}

The author used a generative AI tool solely for language refinement and
clarity improvement.  All scientific content, derivations, analysis, and
conclusions are entirely the responsibility of the author.

\section*{Conflicts of Interest}

The author declares that there is no conflict of interest in this
manuscript.

\section*{Data Availability Statement}

No data were used in this study.




\begin{thebibliography}{99}

\bibitem{Jackiw1976}
  R.~Jackiw, C.~Rebbi,
  Solitons with fermion number $1/2$,
  \textit{Phys.\ Rev.\ D} \textbf{13} (1976) 3398--3409.

\bibitem{CastroNeto2009}
  A.~H.~Castro~Neto, F.~Guinea, N.~M.~R.~Peres,
  K.~S.~Novoselov, A.~K.~Geim,
  The electronic properties of graphene,
  \textit{Rev.\ Mod.\ Phys.} \textbf{81} (2009) 109--162.

\bibitem{Vozmediano2010}
  M.~A.~H.~Vozmediano, M.~I.~Katsnelson, F.~Guinea,
  Gauge fields in graphene,
  \textit{Phys.\ Rep.} \textbf{496} (2010) 109--148.

\bibitem{Chen2023}
  K.~Chen, F.~Komissarenko, D.~Smirnova, A.~Vakulenko,
  S.~Kiriushechkina, I.~Volkovskaya, S.~Guddala, V.~Menon,
  A.~Al\`u, A.~B.~Khanikaev,
  Photonic Dirac cavities with spatially varying mass term,
  \textit{Sci.\ Adv.} \textbf{9} (2023) eabq4243.

\bibitem{Kim2025}
  S.~Kim, K.~Kim,
  Spatial localization and diffusion of Dirac particles and waves induced
  by random temporal medium variations,
  \textit{Commun.\ Phys.} \textbf{8} (2025) 32.
  
  \bibitem{tarantola} Rossi, S., and Tarantola, A., "Topology of 2D Dirac operators with variable mass and an application to shallow-water waves", Journal of Physics A: Mathematical and Theoretical, v. 57, n. 6, (2024) 065201.

\bibitem{BenDaniel1966}
  D.~J.~BenDaniel, C.~B.~Duke,
  Space-charge effects on electron tunneling,
  \textit{Phys.\ Rev.} \textbf{152} (1966) 683--692.

\bibitem{vonRoos1983}
  O.~von~Roos,
  Position-dependent effective masses in semiconductor theory,
  \textit{Phys.\ Rev.\ B} \textbf{27} (1983) 7547--7552.

\bibitem{Bastard1988}
  G.~Bastard,
  \textit{Wave Mechanics Applied to Semiconductor Heterostructures},
  Les Editions de Physique, Les Ulis, 1988.

\bibitem{LiKuhn1993}
  T.~Li, K.~J.~Kuhn,
  Band-offset ratio dependence on the effective-mass Hamiltonian,
  \textit{Phys.\ Rev.\ B} \textbf{47} (1993) 12760--12763.

\bibitem{Mustafa2007}
  O.~Mustafa, S.~H.~Mazharimousavi,
  Interaction of a nonrelativistic particle with a position-dependent mass
  in a generalized Morse field,
  \textit{Ann.\ Phys.} \textbf{322} (2007) 1531--1539.

\bibitem{deSouzaDutra2000}
  A.~de~Souza~Dutra, C.~A.~S.~Almeida,
  Exact solvability of potentials with spatially dependent effective masses,
  \textit{Phys.\ Lett.\ A} \textbf{275} (2000) 25--30.

\bibitem{Znojil2001}
  M.~Znojil, F.~Cannata, B.~Bagchi, R.~Roychoudhury,
  PT-symmetric Hamiltonians with variable mass,
  \textit{Phys.\ Lett.\ B} \textbf{503} (2001) 216--222.

\bibitem{Lima2023}
  R.~M.~Lima, H.~R.~Christiansen,
  The kinetic Hamiltonian with position-dependent mass,
  \textit{Physica E} \textbf{150} (2023) 115688.
 
\bibitem{Cavalcante1997}
  F.~S.~A.~Cavalcante, R.~N.~Costa~Filho, J.~R.~Filho,
  C.~A.~S.~de~Almeida, V.~N.~Freire,
  Form of the quantum kinetic-energy operator with spatially varying
  effective mass,
  \textit{Phys.\ Rev.\ B} \textbf{55} (1997) 1326--1328.

\bibitem{Birrell1982}
  N.~D.~Birrell, P.~C.~W.~Davies,
  \textit{Quantum Fields in Curved Space},
  Cambridge University Press, Cambridge, 1982.

\bibitem{Barcelo2011}
  C.~Barcel\'o, S.~Liberati, M.~Visser,
  Analogue gravity,
  \textit{Living Rev.\ Relativ.} \textbf{14} (2011) 3.

\bibitem{Guinea2010}
  F.~Guinea, M.~I.~Katsnelson, A.~K.~Geim,
  Energy gaps and a zero-field quantum Hall effect in graphene by strain
  engineering,
  \textit{Nat.\ Phys.} \textbf{6} (2010) 30--33.

\bibitem{Levy2010}
  N.~Levy, S.~A.~Burke, K.~L.~Meaker, M.~Panlasigui, A.~Zettl,
  F.~Guinea, A.~H.~C.~Neto, M.~F.~Crommie,
  Strain-induced pseudo-magnetic fields greater than 300~tesla in graphene
  nanobubbles,
  \textit{Science} \textbf{329} (2010) 544--547.

\bibitem{Jackiw2007}
  R.~Jackiw, S.-Y.~Pi,
  Chiral gauge theory for graphene,
  \textit{Phys.\ Rev.\ Lett.} \textbf{98} (2007) 266402.



\end{thebibliography}
\end{document}